\documentclass[apj]{emulateapj}
\usepackage{natbib}
\usepackage{rotating}

\newcommand{\myemail}{maravena@nrao.edu}

\shorttitle{Identification of two $z>3$ SMGs candidates in the COSMOS field}
\shortauthors{M. Aravena}

\begin{document}


\title{Identification of two bright $z>3$ submillimeter galaxy candidates in the COSMOS field$^{\star}$}

\author{M. Aravena\altaffilmark{1}, J. D. Younger\altaffilmark{2,3}, G. G. Fazio\altaffilmark{4},  M. Gurwell\altaffilmark{4}, D. Espada\altaffilmark{4, 5}, F. Bertoldi\altaffilmark{6}, P. Capak\altaffilmark{7}, D. Wilner\altaffilmark{4}}

\altaffiltext{$\star$}{Based on observations obtained with the SMA, which is a joint project between the Smithsonian Astrophysical Observatory and the Academia Sinica Institute of Astronomy and Astrophysics and is funded by the Smithsonian Institution and the Academia Sinica. Also based on observations obtained, within the COSMOS Legacy Survey, with: the Institut de Radioastronomie Millimetrique (IRAM) 30 m telescope, the Caltech Submillimeter Observatory (CSO), the APEX telescope, the {\em Hubble Space Telescope} (HST), the Spitzer Space telescope, the Subaru telescope, the Kitt Peak National Observatory (KPNO), the Cerro Tololo Inter-American Observatory (CTIO), the National Optical Astronomy Observatory (NOAO), the United Kingdom IR telescope (UKIRT) and the Canada-France-Hawaii telescope (CFHT). The National Radio Astronomy Observatory (NRAO) is a facility of the National Science Foundation operated under cooperative agreement by Associated Universities, Inc.}
\altaffiltext{1}{National Radio Astronomy Observatory. 520 Edgemont Road, Charlottesville VA 22903, USA. \myemail}
\altaffiltext{2}{Institute for Advanced Study, Einstein Drive, Princeton, NJ 08544 USA}
\altaffiltext{3}{Hubble Fellow.}
\altaffiltext{4}{Harvard Smithsonian Center for Astrophysics, 60 Garden St., Cambridge, MA 02138, USA}
\altaffiltext{5}{Instituto de Astrof{\'i}sica de Andaluc{\'i}a - CSIC, Apdo. 3004, 18080 Granada, Spain.}
\altaffiltext{6}{Argelander Institut f\"ur Astronomie. Auf dem H\"ugel 71, 53121 Bonn, Germany.}
\altaffiltext{7}{California Institute of Technology, 1200 East California Boulevard, Pasadena, CA 91125, USA}

\begin{abstract}
We present high-resolution interferometric  Submillimeter Array (SMA) imaging at 890$\mu$m ($\sim2\arcsec$ resolution) of two millimeter selected galaxies -- MMJ100015$+$021549 and MMJ100047$+$021021 -- discovered with the Max-Planck Millimeter Bolometer (MAMBO) on the IRAM 30 m telescope and also detected with Bolocam on the CSO, in the COSMOS field. The first source is significantly detected at the $\sim11\sigma$ level, while the second source is  tentatively detected at the $\sim4\sigma$ level, leading to a positional accuracy of $\sim0.2-0.3\arcsec$. MM100015$+$021549 is identified with a faint radio and $K$-band source. MMJ100047$+$021021 shows no radio emission and is tentatively identified with a very faint $K$-band peak which lies at $\sim1.2\arcsec$ from a clumpy optical source. The submillimeter-to-radio flux ratio for MM100015$+$021549 yields a redshift of $\sim4.8$, consistent with the redshift implied by the UV-to-submillimeter photometry, $z\sim3.0-5.0$. We find evidence for warm dust in this source with an infrared luminosity in the range $\sim0.9-2.5\times10^{13}\ L_\sun$, supporting the increasing evidence for a population luminous submillimeter galaxies at $z>3$.  Finally, the lack of photometric data for MMJ100047$+$021021 does not allow us to investigate its properties  in detail, however its submillimeter-to-radio ratio implies $z>3.5$.
\end{abstract}


\keywords{galaxies: evolution --- galaxies: high-redshift --- galaxies: starburst }



\section{Introduction}

Submillimeter (submm) wavelength blank-field surveys discovered a population of heavily dust-obscured starburst galaxies at high redshifts \citep[e.g.][]{Smail1997,Barger1998,Hughes1998}. These submm galaxies (SMGs) contribute a large fraction of the comoving infrared (IR) luminosity density at high-redshift  \citep{LeFloch2005}, and their clustering properties indicate they could be progenitors of the most massive galaxies at $z<1$ \citep{Blain2004, Viero2009, Aravena2010}. The number counts and dust production rates of these massive, high-redshift starburst galaxies actually place tight constrains on galaxy formation models \citep{Baugh2005}.

The identification of optical counterparts to the submm sources has mainly been based on the identification of radio counterparts, detected through deep Very Large Array (VLA) 20 cm imaging. This technique exploits the local far-IR/radio correlation \citep{Condon1992} and statistical arguments to claim an association between a closeby radio source and the submm source, localizing the SMG with an accuracy of $\sim1\arcsec$. The direct search for counterparts in optical wavebands is not practical due to their common faintness  and the high number density of optical sources within the positional uncertainty. For about $70-80\%$ of all submm sources precise positions can be determined from an identification of faint radio counterparts \citep{Ivison2007}. 

However, the depth of current radio observations limits the identification of submm sources to $z<3$, since due to the strong K-correction, the radio flux (unlike the submm flux) drops rapidly with redshift \citep{Carilli1999}. The only unambiguous way to localize the highest-redshift SMGs in the optical/near-IR is via submm interferometry. 

Radio-selected SMGs have been found to have a median redshift of $z=2.3$ \citep{Chapman2005}. However, recent interferometric submm imaging of a flux-limited sample of SMGs has provided substantial evidence that up to $\sim30\%$ of this population is likely located at $z>3$ \citep{Iono2006,Younger2007,Younger2009}. These radio-dim SMGs appear to be inconspicuous at optical wavelengths, likely obscured by large amounts of dust, and their nature can only be revealed in the IR \citep{Wang2007, Dannerbauer2008, Cowie2009}.  To date, only five such high-redshift SMGs have been confirmed spectroscopically \citep{Capak2008, Schinnerer2008, Coppin2009, Daddi2009a, Daddi2009b, Knudsen2010}.  


In this paper, we report accurate astrometry of two bright SMGs in the COSMOS field whose radio to submm flux ratio indicates that are likely at $z>3$.  We assume a standard cosmology with $H_0=71$ km s$^{-1}$ Mpc$^{-1}$, $\Omega_\Lambda=0.73$ and $\Omega_\mathrm{M}=0.27$.


\section{Observations}

\begin{table}
\centering
\caption{Multi-wavelength photometry$^{a}$\label{table:1}}
\begin{tabular}{lccccc}
\hline 
Band & $\lambda^b$   & MM1 & MM14S & MM14N & Units  \\
\hline \hline
$B^+$  & 0.45    &   $<21$            &   $<21$      &  $24\pm7$      &  nJy\\
$i^+$ & 0.77   &   $56\pm16$     &   $<35$     &  $116\pm11$                  & nJy \\
$K_\mathrm{S}$  & 2.2   &    $768\pm77$      &    $222\pm160$    &  $340\pm140$     & nJy\\
IRAC  & 3.6     &   $1.3\pm0.2$                  &    $<0.2$    &  $0.20\pm0.06$    &  $\mu$Jy\\
IRAC & 4.5    &   $1.9\pm0.3$                  &      $<0.3$  &    $0.32\pm0.11$  & $\mu$Jy \\
MIPS  & 24   &   \ldots                  &   $<0.06$     &  $<0.06$     &  mJy\\
MIPS & 70    &    $<2.1$                &   $<2.7$     &  $<2.7$    & mJy \\
MIPS & 160     &    $<15$              & $<15$       & $<15$    &   mJy \\
LABOCA & 870     &    $16.4\pm1.8$  &  $<10.5$      &  \ldots   &  mJy \\
SMA & 890   &   $16.8\pm1.5$           &  $8.5\pm2.0$      & $<6$    &  mJy \\
Bolocam &1100    &    $6.9\pm1.9$             &   $3.8\pm19$     & \ldots     & mJy \\
MAMBO & 1200    &    $6.2\pm0.9$              &   $4.1\pm1.0$     & \ldots     &  mJy\\
VLA & 20   &   $35\pm10$          & $<30$   &  $<30$     &   $\mu$Jy\\
\hline
\end{tabular}
\begin{flushleft}
\indent $^{a}$ Measurements at $3.6$ and $4.5\ \mu$m based on the deep {\it Spitzer} warm mission exposures.
\indent $^b$ Wavelength in microns, except for the VLA where it is given in centimeters.
\indent $^c$ Photometry in the MIPS 24 $\mu$m bands not possible due to blending with a bright source. 
\end{flushleft}
\end{table}

\subsection{The COSMOS field}

COSMOS is the largest deep survey carried out with the {\it Hubble Space telescope} (HST) covering $\sim2$ deg$^{2}$ in the sky. Extensive imaging of the COSMOS field has been performed from the X-rays to the radio wavelengths. This includes: complete optical/near-IR coverage in 22 broad and intermediate bands with several ground based observatories including the Subaru telescope, the CFHT, the UKIRT and the KPNO; IR imaging with {\em Spitzer}, including new deep 3.6 and 4.5 $\mu$m images obtained as part of its warm mission; and radio imaging with the VLA at 20 cm. For details in the optical/IR imaging and catalogs of the COSMOS field see \citet{Capak2007} and \citet{Ilbert2009}. A complete description of the radio imaging is given in \citet{Schinnerer2007}. 

\subsection{Sample selection}

In the course of the MAMBO 1.2 mm survey of the COSMOS field \citep{Bertoldi2007}, fifteen sources were detected with $\mathrm{S/N}> 4$, five of which do not have a significant radio counterpart ( $< 30\ \mu$Jy, 3$\sigma$). From these five radio-faint mm sources, we selected two that were also significantly detected at 1.1 mm with Bolocam: MMJ100016$+$021549  and MMJ100047$+$021018. Hereafter, we refer to these sources as MM1 and MM14, respectively, following \citet{Bertoldi2007}. The deboosted flux densities for MM1 were $S_\mathrm{1.2 mm}=6.2\pm0.9$ mJy and $S_\mathrm{1.1 mm}=5.9\pm1.9$ mJy in the MAMBO and Bolocam maps, respectively \citep{Bertoldi2007}. This source was recently detected with the Large Bolometer Camera (LABOCA) with $S_{870 \mu \mathrm{m}}=16.4\pm1.8$ mJy (Albrecht et al., in preparation). The MAMBO and Bolocam deboosted flux densities for MM14 were $S_\mathrm{1.2 mm}=4.1\pm1.0$ mJy and $S_\mathrm{1.1 mm}=3.6\pm1.9$ mJy, respectively \citep{Bertoldi2007}, however no significant emission is seen in the LABOCA image down to 10.5 mJy ($\sim$3$\sigma$).



\begin{figure}
\centering
\includegraphics[scale=0.32]{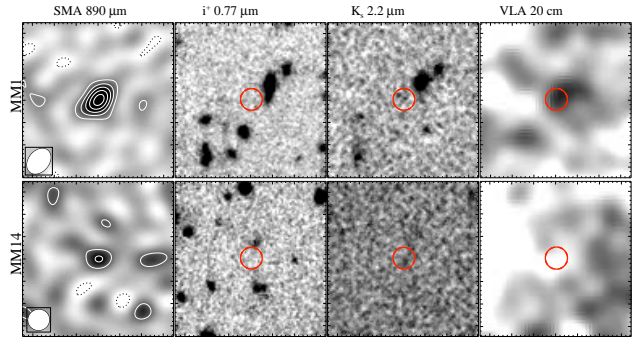}
\caption{Postage stamps centered at the position of the SMA detections. Images are $14\arcsec\times14\arcsec$ in size. The images to the left show the SMA 890 $\mu$m emission at significance levels of -4, -2, 2, 4, 6, 8 and 10$\sigma$ (see text). Solid and dashed contours represent positive and negative fluxes, respectively. The red 2\arcsec \ diameter circle represents the position of the SMA detection.\label{fig:allphot}}
\end{figure}

\subsection{SMA 890 $\mu$m imaging}

SMA observations of MM1 and MM14 were carried out in 2009 January 10 and 2009 May 07, respectively, under good weather conditions ($\tau_{250 \mathrm{GHz}}<0.1$). The receivers have two sidebands, each with 2 GHz bandwidth, which when averaged yield a 4 GHz effective bandwidth centered at $345$ GHz ($\lambda\sim890\ \mu$m). Seven antennas, arranged in the compact (COM) configuration\footnote{half-power beam width (HPBW) $\approx2\arcsec$; see {\it http://sma1.sma.hawaii.edu/specs.html} for more information.}  with the MAMBO source positions as phase tracking centers. 

The data were calibrated with the MIR package \citep{Scoville1993} specially adapted for SMA data, using the strong continuum source 3C273  ($S_{345\mathrm{GHz}}\sim4.9$ Jy) as passband calibrator and Ceres ($S_{345\mathrm{GHz}}\sim4.2$ Jy) for primary flux calibration. The flux scale is estimated to be accurate within 20\%. The quasars J0854$+$201 ($\sim$3.1 Jy;  $\sim24.2\degr$ away) and J1058$+$015 ($\sim1.7$ Jy; $\sim14.5\degr$ away) were observed every $\sim25$ min for gain calibration.  Following \citet{Younger2008, Younger2009, Younger2010}, we also performed hourly scans of a dimmer, but significantly closer test quasar J1008$+$063 ($\sim0.11$ Jy; $\sim4.5\degr$ away) to empirically verify the phase transfer and estimate the positional uncertainty. The visibility data for MM1 showed good phase stability, however, about half of the data for MM14 had to be flagged due to bad phases. 

The calibrated visibility data were imaged using the AIPS software.  We used the AIPS task IMAGR, which uses the CLEAN algorithm, and natural weighting to deconvolve the images down to $1\sigma$ in a box centered on our targets. This led to beam sizes of $2.55\arcsec\times1.86\arcsec$ and $1.96\arcsec\times1.86\arcsec$ and noise levels of $1.5$ mJy and $1.95$ mJy beam$^{-1}$ for MM1 and MM14, respectively. Finally, fluxes were measured with Gaussian fits using the JMFIT task, included in the AIPS software.

\section{Results and Analysis}

\begin{figure}
\centering
\includegraphics[scale=0.4]{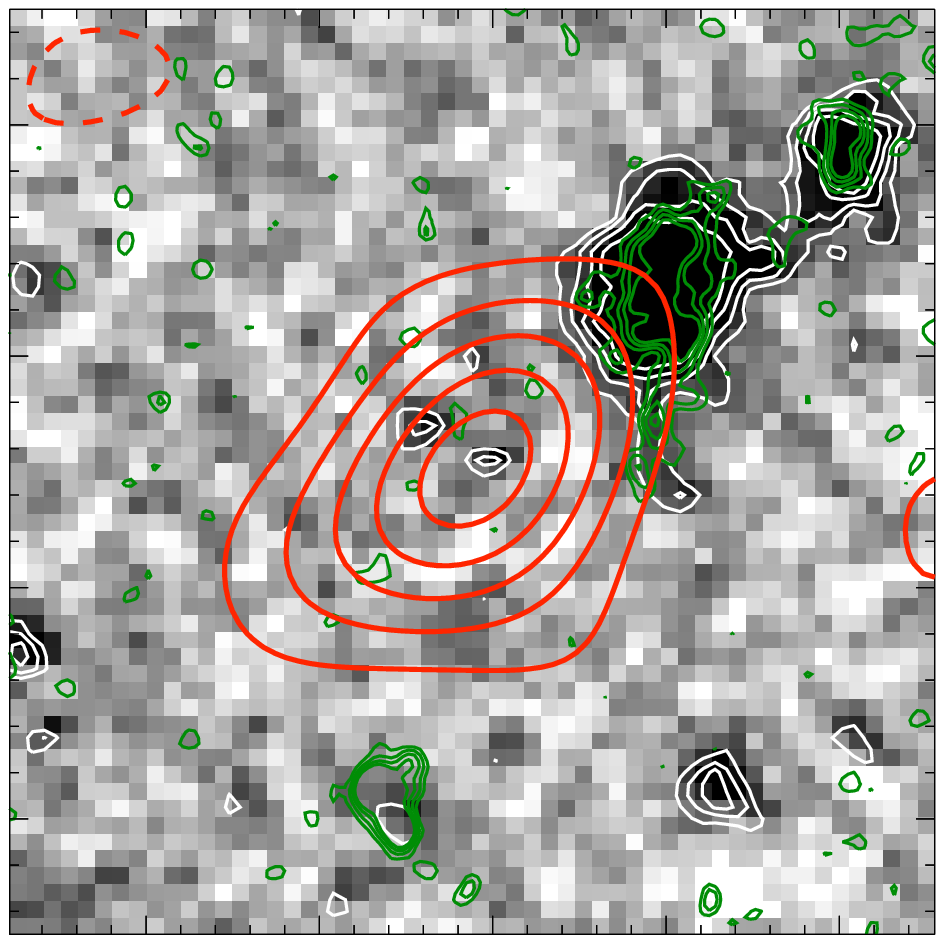}
\includegraphics[scale=0.4]{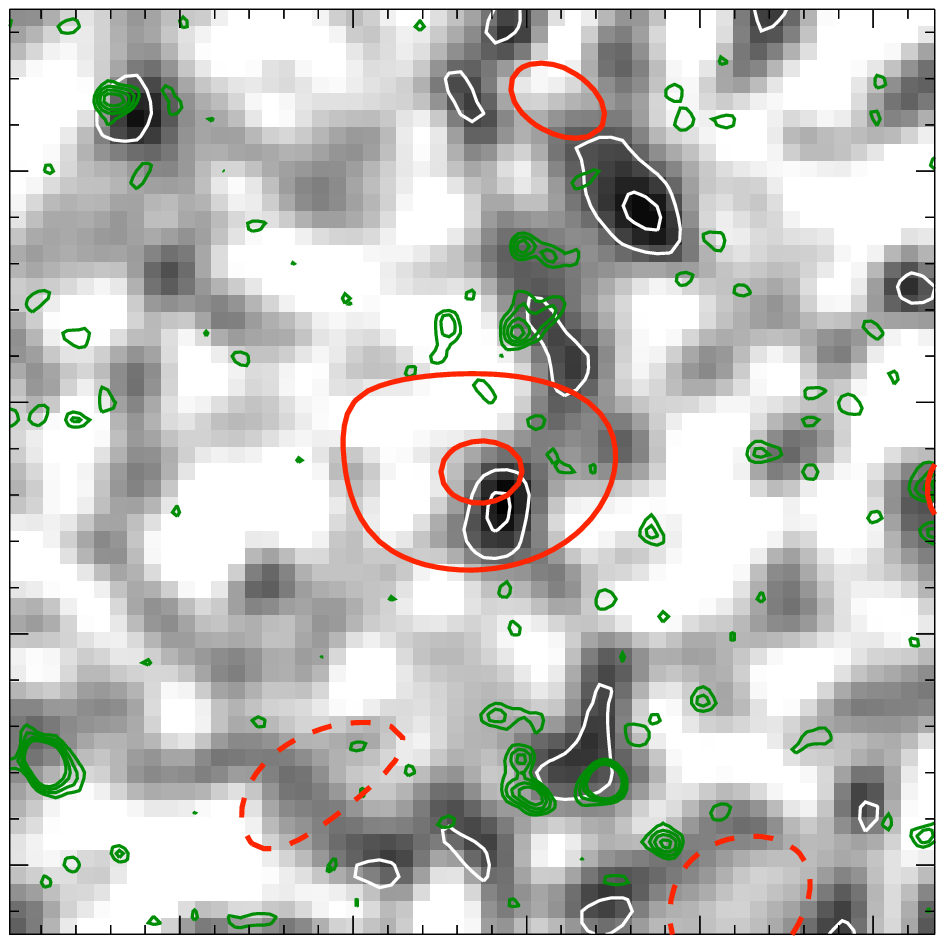}
\caption{$K$-band $8\arcsec\times8\arcsec$ postage stamp images centered at the position of the SMA detections of MM1 (left) and MM14 (right), with $K$-band white contours overlaid at 2, 3, 4 and 5$\sigma$ levels. Green contours represent the HST F814W image smoothed with a Gaussian kernel of 3 pixels (0.15\arcsec) at 2, 3, 4 and 5$\sigma$ levels. The red contours represent the 890 $\mu$m emission with levels as in Fig. \ref{fig:allphot}. The $K$-band image for MM14 has also been smoothed with a Gaussian kernel of 3 pixels (0.45\arcsec) to enhance the significance level. \label{fig:closeup}}
\end{figure}

\subsection{SMA detections}

Fitting two-dimensional Gaussians to the SMA images, leads to flux densities of  $16.8\pm1.5$ mJy and  $8.5\pm2.0$ mJy for MM1 and MM14, respectively. The MM14 detection is tentative since, as we shall see below, it could not be reliably identified with any significant counterpart at other wavelengths. 

The Gaussian fit indicates that both sources are unresolved with a maximum deconvolved FWHM size of $1.8\arcsec\times1.0\arcsec$ and $2.0\arcsec\times0.6\arcsec$, respectively, consistent with the sizes found for high-redshift SMGs, which are typically unresolved at $\sim2\arcsec$ resolution \citep{Iono2006,Younger2007, Wang2007,Younger2009}. This also yields from the flatness of the real visibility amplitudes as a function of the projected baseline length. The measured SMA position for MM1 is $\alpha(\mathrm{J2000})=10^\mathrm{h}00^\mathrm{m}15.612^\mathrm{s}$, $\delta(\mathrm{J2000})=+02\degr15\arcmin49.00\arcsec$, with a positional error in the Gaussian fit of $0.09\arcsec$, while for MM14 it is $\alpha(\mathrm{J2000})=10^\mathrm{h}00^\mathrm{m}47.329^\mathrm{s}$, $\delta(\mathrm{J2000})=+02\degr10\arcmin21.44\arcsec$, with a positional error in the fit of $0.15\arcsec$. This positional accuracy in the fit is consistent with the one expected for the beam and S/N of the observations: 0.09\arcsec and 0.2\arcsec for MM1 and MM14, respectively. From the comparison of the reference \citep{Browne1998} and measured (this work) positions of the test quasar J1008$+$063, we find a positional uncertainty of $0.18\arcsec$. This, added in quadrature to the positional error in the Gaussian fit to the MM1 and MM14 images, gives a positional uncertainty of $0.2\arcsec$ and $0.27\arcsec$, respectively.

\subsection{Multi-wavelength counterparts}

{\it MM1}. The SMA peak position coincides with the position of a $\sim3.5\sigma$ radio peak (Fig. \ref{fig:allphot}). A bright and elongated source with photometric redshift $z=1.4$ \citep{Ilbert2009} is located at $\sim$2.1\arcsec north-west from the SMA position. The radio peak lies within $0.3\arcsec$ from the SMA position, however given the beam of the radio image ($\sim2\arcsec$), we do not discard that part of the emission comes from this bright optical source. From Fig. \ref{fig:closeup} (left), we identify a very faint $K$-band source, located at $\approx0.3\arcsec$ from the SMA position, as the likely counterpart.  The bright optical source strongly contaminates the {\it Spitzer} images, making it difficult to reliably measure the faint emission of MM1. We extracted photometry in the IRAC bands by subtracting this bright source based on the K-band image convolved with the IRAC PSF (Table \ref{table:1}). We do not attempt to extract photometry of this source in the 24 $\mu$m images. No emission is detected at 70 and 160 $\mu$m. 

The deep $K$-band images show two peaks, separated by about 0.6\arcsec, or a physical scale of $\sim4.3$ kpc at $z\sim3$ (Fig. \ref{fig:closeup}). The fainter peak appears to be the one associated with the submm emission, suggesting a possible double system, similar to the case of the SMG AzTEC11 \citep{Younger2009}.

{\it MM14}. The radio maps do not show any peak close to the position of the SMA source down to a $3\sigma$ level of 30 $\mu$Jy.  At $\sim1.2\arcsec$ to the north of the SMA position we find a faint optical source that appears diffuse and faint in the $K$-band (Fig. \ref{fig:allphot}). This source has a likely photometric redshift of $\sim3.4$ \citep{Mobasher2007}. 

From Fig. \ref{fig:allphot}, the northern optical source appears to be composed by several ``clumps''  that extend over $\sim1.5\arcsec$, or $\sim11$ kpc at $z\sim3.5$. We also find a very faint $K$-band emission peak ($\sim$3$\sigma$ in the smoothed image), located 0.4\arcsec \ to the south of the SMA position (Fig. \ref{fig:closeup}). This peak has $\sim2\sigma$ significance in the original $K$-band image (without smoothing; Fig. \ref{fig:allphot}), implying the source is spatially extended. Due to its proximity to the SMA position, it appears to be the most likely counterpart to the submm emission despite its faintness. Hereafter, we refer to this source as MM14S (south), while for the northern optical source we refer as MM14N. However, since we could not reliably identify any significant multi-wavelength counterpart for this source and given the relatively low significance of the SMA detection, we label this source (MM14S) as a tentative detection.

The spatial configuration between MM14S and MM14N could resemble a merger/interaction system, where the submm emission comes from a highly obscured source (MM14S), but could also correspond to an extended galaxy with the submm emission located in an obscured spiral arm. The offset between the MM14N and the submm position (MM14S) is $\sim9$ kpc, assuming $z\sim3.5$, similar to the case of the high-redshift SMG GN20 \citep{Iono2006}, where the submm and the optical peaks are separated by $\sim0.8\arcsec$, or $\sim6$ kpc. Based on the local density of sources with $z>3$, $n=0.002$ arcsec$^{-2}$, we find that the probability that a $z>3$ optical source brighter than MM14N is located by chance within a distance of 1.2\arcsec \ from the SMA position is only $P=0.9\%$, thus supporting a physical association.

Optical/IR photometry was performed with SExtractor \citep{Bertin1996} in a 2\arcsec \ aperture, using the $K$-band images for detection. None of our targets was detected in the {\it Spitzer} MIPS bands, and we thus provide $3\sigma$ upper limits based on their local noise level within one beam. The measured flux densities at several wavelengths for MM1, MM14S and MM14N are listed in Table \ref{table:1}.

\subsection{Photometric redshifts}

\begin{figure}
\centering
\includegraphics[scale=0.45]{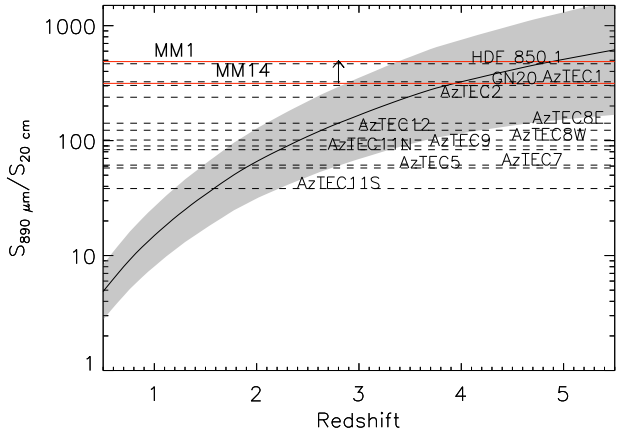}
\caption{Redshift estimation based on the submm-to-radio flux ratio. The solid line represents this ratio computed for the IR luminous starburst galaxy Arp220 (which roughly has $T_\mathrm{dust}=45$ K). The filled area shows the flux ratio obtained for similar models with dust temperatures from 30 to 60 K. Dashed horizontal lines show the flux ratios measured for the radio-identified submm sources with SMA detections \citep{Younger2007, Younger2009}, and for the high-redshift SMGs HDF 850.1 and GN20 \citep{Cowie2009, Iono2006}. The upper pointing arrow represents the lower limit in the flux ratio for MM14S.\label{fig:fluxratio}}
\end{figure}

MM1 and MM14S were not detected in the COSMOS catalogs \citep{Capak2007, Ilbert2009} and thus there is no previous estimate for their redshift. 

Assuming that the correlation between the far-IR and radio integrated emission from local starburst galaxies also holds for distant galaxies, the submm-to-radio flux ratio can be used as a measure of redshift \citep{Carilli1999}. Figure \ref{fig:fluxratio} shows the  890 $\mu$m to 20 cm flux ratio as a function of redshift for the prototypical starburst galaxy Arp220, and illustrates the uncertainties introduced by different dust temperatures. From this, we derive a redshift of 4.8$_{-1.4}^{+2.5}$ for MM1, and a lower limit $z>3.5$ for MM14S. 

Given the lack of optical/IR photometric information for MM14S, we do not attempt to derive the redshift and dust properties of this source, and hereafter we focus on MM1 and MM14N.

We derived optical/IR photometric redshifts for MM1 and MM14N using the {\it HyperZ} code \citep{Bolzonella2000}. We allowed for a redshift range $z=0-7$, and using a \citet{Calzetti2000} extinction law with an extinction range $A_\mathrm{V}=0-4$. For comparison, we used two different set of templates libraries:  5 solar metallicity stellar population models from \citet{bc2003}, similar to the {\it HyperZ} defaults; and 10 templates from the SWIRE library \citep{Polletta2007}, including models for an elliptical, three spiral, four starburst and three AGN dominated galaxies.

As shown in Fig. \ref{fig:sed}, the $\chi^2$ distribution and redshift solutions for MM1 and MM14N obtained using both set of templates are similar. Using the \citet{bc2003} templates, the best fit for MM1 is produced at $z=3.1_{-0.6}^{+0.5}$ by a spiral galaxy model, with $A_\mathrm{V}=1.4$ (black curve); while the best fit for MM14N is produced at $z=3.7_{-0.3}^{+0.3}$ by a single burst model, with $A_\mathrm{V}=1.0$. Using the SWIRE library, the best fits are produced by a late spiral galaxy at $z=3.0_{-0.9}^{+0.5}$ and a QSO template at $z=3.6_{-0.3}^{+0.2}$ for MM1 and MM14N, respectively. The quoted uncertainties correspond to the 90\% confidence level derived from the fitting routine.

Each of the \citet{bc2003} templates follow a prescription for the stellar mass to luminosity ratio (e.g. as a function of age). Using the best fits and assuming a \citet{Chabrier2003} initial mass function, we find stellar masses of $\sim1.0\times10^{11}\ M_\sun$ and $\sim3.6\times10^{10}\ M_\sun$, for MM1 and MM14N, respectively. We note, however, that these stellar mass estimates are rough given the large uncertainty in redshift. 

For MM1, we also computed redshifts by fitting templates to the UV-to-mm photometry. We used the same SWIRE templates, redshift and extinction ranges. The best fit is produced by an Arp220 model, with $A_\mathrm{V}=0.3$ and $z=4.5$. The value of $\chi^2$ is dominated by the ratio between the submm to optical light, indicating a possible range of redshift from $\sim$3.0 to 5.0. Our redshift estimates are in good agreement with the redshift implied by the submm to radio flux ratio and supports a higher redshift than that obtained purely from the optical photometry.

\subsection{Dust properties}

\begin{figure}
\centering
\includegraphics[scale=0.35]{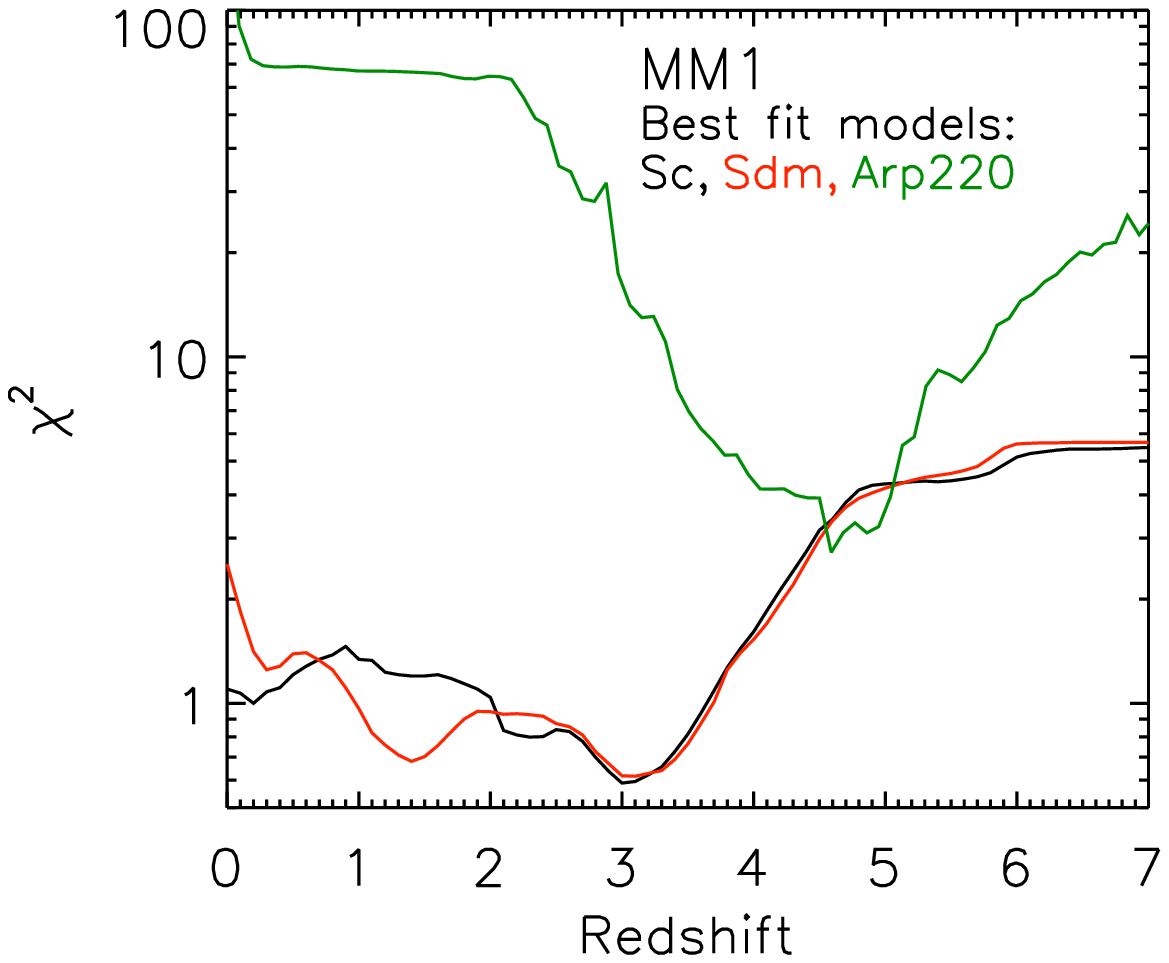}
\includegraphics[scale=0.35]{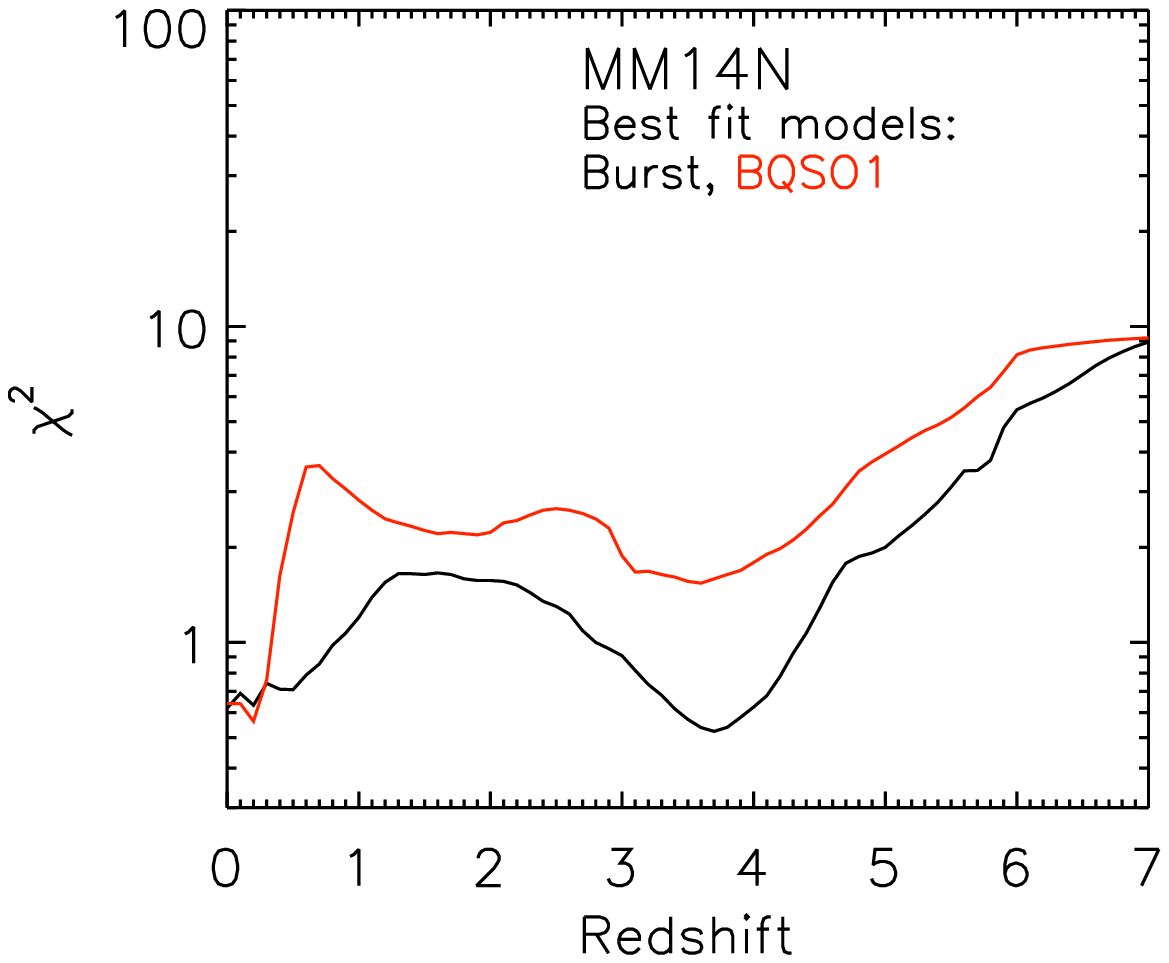}
\includegraphics[scale=0.35]{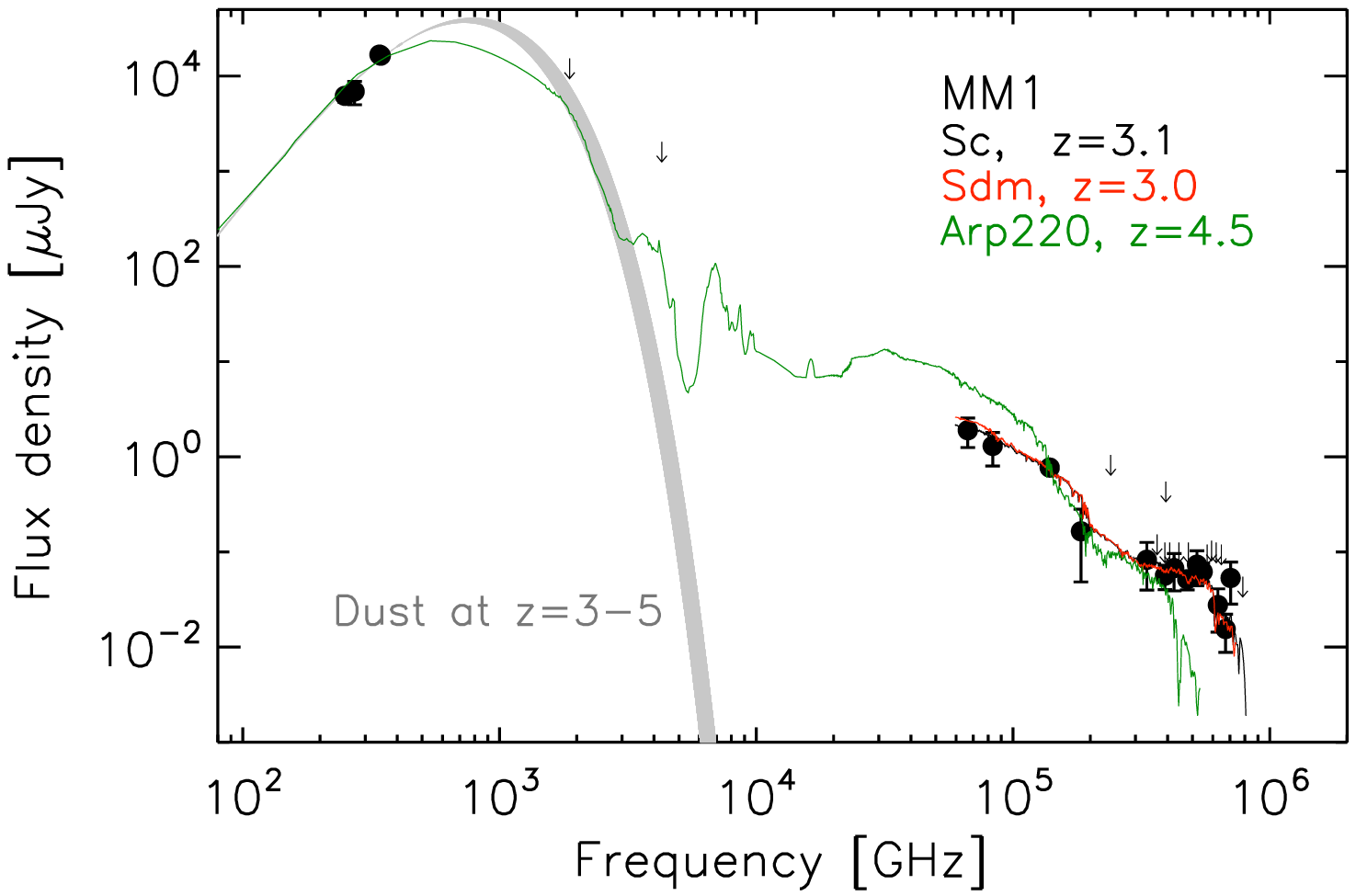}
\caption{Photometric redshift results.  {\it Top:} the $\chi^2$ values as a function of redshift obtained by fitting the optical photometry with the the \citet{bc2003} templates (black curves) and the SWIRE templates (red curves) for MM1 (left) and MM14N (right); and by fitting the UV-to-mm photometry with the SWIRE templates (green curves) for MM1.  {\it Bottom:}  SED of MM1. The colors represent the different template libraries used.  The gray curve shows the best fit dust model to the far-IR data at $z=3-5$.\label{fig:sed}}
\end{figure}

As illustrated in Fig. \ref{fig:sed}, the far-IR SED of MM1 can be described by a modified black body spectrum. We fitted a model of the form $S_\nu\propto B_\nu (T_\mathrm{dust}) (1-e^{-\tau_\nu})$, where $B_\nu (T_\mathrm{dust})$ is the Planck function. The opacity, $\tau_\nu$, is proportional to the dust mass $M_\mathrm{dust}$ and to $\nu^\beta$, with the emissivity index $\beta=1.5$. Given the large uncertainties in redshift, we consider a range of redshift $z=3-5$ to model the photometric data ($70-1200\mu$m). 

We find dust temperatures and masses in the range $T_\mathrm{dust}=50-65$ K and $M_\mathrm{dust}=1.2-2.6\times10^{9}\ M_\sun$.  These models imply far-IR luminosities $L_{8-1000\mu\mathrm{m}}=0.9-2.5\times10^{13}\ L_\sun$. Using the computed IR luminosities and the \citet{Kennicutt1998} empirical calibration, we compute star-formation rates (SFRs) of $1800-5000\ M_\sun$ yr$^{-1}$ for this source. 

\section{Discussion}

Our detection of two out of the five significant radio-faint sources discovered in the MAMBO survey of the COSMOS field supports the increasing evidence for a population of luminous SMGs at $z>3$. It has recently been suggested that these galaxies may be the progenitors of quiescent, massive galaxies at $z\sim2$ \citep{Capak2008, Coppin2009, Coppin2010}. The old stellar populations found in these $z\sim2$ quiescent galaxies require to have formed at $z>3$ in short bursts of star formation. 

If we assume that four of these five MAMBO sources with no radio significant radio counterparts are real and located at $z>3$ (e.g., we expect that one of the five is a fake detection due to flux boosting), it implies that up to $\sim30\%$ (4/15) of the $S_\mathrm{250 GHz}>4$ mJy detections are located at $z>3$. This is well justified by the confirmation of MM1 and tentative detection of MM14 with the SMA, and of two other MAMBO sources with LABOCA (Albrecht et al., in preparation). Assuming that $2-4$ of the MAMBO sources are uniformly distributed over the comoving volume spanned by $z=3-5$ within the area of the MAMBO survey, we find a surface density of $\sim13-30$ deg$^{-2}$ and a volume density of $\sim1-6\times10^{-6}$ Mpc$^{-3}$. This is consistent with the upper limit of $>10^{-6}$ Mpc$^{-3}$ ($>6$ deg$^{-2}$) from the spectroscopically confirmed $z>4$ SMGs \citep{Coppin2009} and with the density of $z\sim3-5$ systems with baryonic masses $>10^{11}\ M_\sun$ from predictions of theoretical models \citep{Baugh2005}.


\acknowledgments
JDY acknowledges support from NASA through Hubble Fellowship grant \#HF-51266.01 awarded by the Space Telescope Science Institute,which
is operated by the Association of Universities for Research in Astronomy, Inc., for NASA, under contract NAS 5-26555. DE was supported by a Marie Curie International Fellowship within the 6$\rm ^{th}$ European Community Framework Programme (MOIF-CT-2006-40298).




\begin{thebibliography}{48}
\expandafter\ifx\csname natexlab\endcsname\relax\def\natexlab#1{#1}\fi

\bibitem[{{Aravena} {et~al.}(2010){Aravena}, {Bertoldi}, {Carilli},
  {Schinnerer}, {McCracken}, {Salvato}, {Riechers}, {Sheth}, {Sm{\v
  o}lci{\'c}}, {Capak}, {Koekemoer}, \& {Menten}}]{Aravena2010}
{Aravena}, M., {et~al.} 2010, \apjl, 708, L36

\bibitem[{{Barger} {et~al.}(1998){Barger}, {Cowie}, {Sanders}, {Fulton},
  {Taniguchi}, {Sato}, {Kawara}, \& {Okuda}}]{Barger1998}
{Barger}, A.~J., {Cowie}, L.~L., {Sanders}, D.~B., {Fulton}, E., {Taniguchi},
  Y., {Sato}, Y., {Kawara}, K., \& {Okuda}, H. 1998, \nat, 394, 248

\bibitem[{{Baugh} {et~al.}(2005){Baugh}, {Lacey}, {Frenk}, {Granato}, {Silva},
  {Bressan}, {Benson}, \& {Cole}}]{Baugh2005}
{Baugh}, C.~M., {Lacey}, C.~G., {Frenk}, C.~S., {Granato}, G.~L., {Silva}, L.,
  {Bressan}, A., {Benson}, A.~J., \& {Cole}, S. 2005, \mnras, 356, 1191

\bibitem[{{Bertin} \& {Arnouts}(1996)}]{Bertin1996}
{Bertin}, E. \& {Arnouts}, S. 1996, \aaps, 117, 393


\bibitem[{{Bertoldi} {et~al.}(2007){Bertoldi}, {Carilli}, {Aravena},
  {Schinnerer}, {Voss}, {Smolcic}, {Jahnke}, {Scoville}, {Blain}, {Menten},
  {Lutz}, {Brusa}, {Taniguchi}, {Capak}, {Mobasher}, {Lilly}, {Thompson},
  {Aussel}, {Kreysa}, {Hasinger}, {Aguirre}, {Schlaerth}, \&
  {Koekemoer}}]{Bertoldi2007}
{Bertoldi}, F., {et~al.} 2007, \apjs, 172, 132

\bibitem[{{Blain} {et~al.}(2004){Blain}, {Chapman}, {Smail}, \&
  {Ivison}}]{Blain2004}
{Blain}, A.~W., {Chapman}, S.~C., {Smail}, I., \& {Ivison}, R. 2004, \apj, 611,
  725

\bibitem[{{Bolzonella} {et~al.}(2000){Bolzonella}, {Miralles}, \&
  {Pell{\'o}}}]{Bolzonella2000}
{Bolzonella}, M., {Miralles}, J., \& {Pell{\'o}}, R. 2000, \aap, 363, 476


\bibitem[{{Browne} {et~al.}(1998){Browne}, {Wilkinson}, {Patnaik}, \&
  {Wrobel}}]{Browne1998}
{Browne}, I.~W.~A., {Wilkinson}, P.~N., {Patnaik}, A.~R., \& {Wrobel}, J.~M.
  1998, \mnras, 293, 257

\bibitem[{{Bruzual} \& {Charlot}(2003)}]{bc2003}
{Bruzual}, G., \& {Charlot}, S. 2003, \mnras, 344, 1000

\bibitem[{{Calzetti} {et~al.}(2000){Calzetti}, {Armus}, {Bohlin}, {Kinney},
  {Koornneef}, \& {Storchi-Bergmann}}]{Calzetti2000}
{Calzetti}, D., {Armus}, L., {Bohlin}, R.~C., {Kinney}, A.~L., {Koornneef}, J.,
  \& {Storchi-Bergmann}, T. 2000, \apj, 533, 682

\bibitem[{{Capak} {et~al.}(2007){Capak}, {Aussel}, {Ajiki}, {McCracken},
  {Mobasher}, {Scoville}, {Shopbell}, {Taniguchi}, {Thompson}, {Tribiano},
  {Sasaki}, {Blain}, {Brusa}, {Carilli}, {Comastri}, {Carollo}, {Cassata},
  {Colbert}, {Ellis}, {Elvis}, {Giavalisco}, {Green}, {Guzzo}, {Hasinger},
  {Ilbert}, {Impey}, {Jahnke}, {Kartaltepe}, {Kneib}, {Koda}, {Koekemoer},
  {Komiyama}, {Leauthaud}, {Le Fevre}, {Lilly}, {Liu}, {Massey}, {Miyazaki},
  {Murayama}, {Nagao}, {Peacock}, {Pickles}, {Porciani}, {Renzini}, {Rhodes},
  {Rich}, {Salvato}, {Sanders}, {Scarlata}, {Schiminovich}, {Schinnerer},
  {Scodeggio}, {Sheth}, {Shioya}, {Tasca}, {Taylor}, {Yan}, \&
  {Zamorani}}]{Capak2007}
{Capak}, P., {et~al.} 2007, \apjs, 172, 99

\bibitem[{{Capak} {et~al.}(2008){Capak}, {Carilli}, {Lee}, {Aldcroft},
  {Aussel}, {Schinnerer}, {Wilson}, {Yun}, {Blain}, {Giavalisco}, {Ilbert},
  {Kartaltepe}, {Lee}, {McCracken}, {Mobasher}, {Salvato}, {Sasaki}, {Scott},
  {Sheth}, {Shioya}, {Thompson}, {Elvis}, {Sanders}, {Scoville}, \&
  {Tanaguchi}}]{Capak2008}
---. 2008, \apjl, 681, L53

\bibitem[{{Carilli} \& {Yun}(1999)}]{Carilli1999}
{Carilli}, C.~L., \& {Yun}, M.~S. 1999, \apjl, 513, L13


\bibitem[{{Chabrier}(2003)}]{Chabrier2003}
{Chabrier}, G. 2003, \apjl, 586, L133

\bibitem[{{Chapman} {et~al.}(2005){Chapman}, {Blain}, {Smail}, \&
  {Ivison}}]{Chapman2005}
{Chapman}, S.~C., {Blain}, A.~W., {Smail}, I., \& {Ivison}, R.~J. 2005, \apj,
  622, 772

\bibitem[{{Condon}(1992)}]{Condon1992}
{Condon}, J.~J. 1992, \araa, 30, 575

\bibitem[{{Coppin} {et~al.}(2010){Coppin}, {Chapman}, {Smail}, {Swinbank},
  {Walter}, {Wardlow}, {Weiss}, {Alexander}, {Brandt}, {Dannerbauer}, {De
  Breuck}, {Dickinson}, {Dunlop}, {Edge}, {Emonts}, {Greve}, {Huynh}, {Ivison},
  {Knudsen}, {Menten}, {Schinnerer}, \& {van der Werf}}]{Coppin2010}
{Coppin}, K., {et~al.} 2010, ArXiv:1004.4001


\bibitem[{{Coppin} {et~al.}(2009){Coppin}, {Smail}, {Alexander}, {Weiss},
  {Walter}, {Swinbank}, {Greve}, {Kovacs}, {De Breuck}, {Dickinson}, {Ibar},
  {Ivison}, {Reddy}, {Spinrad}, {Stern}, {Brandt}, {Chapman}, {Dannerbauer},
  {van Dokkum}, {Dunlop}, {Frayer}, {Gawiser}, {Geach}, {Huynh}, {Knudsen},
  {Koekemoer}, {Lehmer}, {Menten}, {Papovich}, {Rix}, {Schinnerer}, {Wardlow},
  \& {van der Werf}}]{Coppin2009}
---. 2009, \mnras, 395, 1905

\bibitem[{{Cowie} {et~al.}(2009){Cowie}, {Barger}, {Wang}, \&
  {Williams}}]{Cowie2009}
{Cowie}, L.~L., {Barger}, A.~J., {Wang}, W., \& {Williams}, J.~P. 2009, \apjl,
  697, L122

\bibitem[{{Daddi} {et~al.}(2009{\natexlab{a}}){Daddi}, {Dannerbauer}, {Krips},
  {Walter}, {Dickinson}, {Elbaz}, \& {Morrison}}]{Daddi2009b}
{Daddi}, E., {Dannerbauer}, H., {Krips}, M., {Walter}, F., {Dickinson}, M.,
  {Elbaz}, D., \& {Morrison}, G.~E. 2009{\natexlab{a}}, \apjl, 695, L176

\bibitem[{{Daddi} {et~al.}(2009{\natexlab{b}}){Daddi}, {Dannerbauer}, {Stern},
  {Dickinson}, {Morrison}, {Elbaz}, {Giavalisco}, {Mancini}, {Pope}, \&
  {Spinrad}}]{Daddi2009a}
{Daddi}, E., {et~al.} 2009{\natexlab{b}}, \apj, 694, 1517

\bibitem[{{Dannerbauer} {et~al.}(2008){Dannerbauer}, {Walter}, \&
  {Morrison}}]{Dannerbauer2008}
{Dannerbauer}, H., {Walter}, F., \& {Morrison}, G. 2008, \apjl, 673, L127

\bibitem[{{Greve} {et~al.}(2005){Greve}, {Bertoldi}, {Smail}, {Neri},
  {Chapman}, {Blain}, {Ivison}, {Genzel}, {Omont}, {Cox}, {Tacconi}, \&
  {Kneib}}]{Greve2005}
{Greve}, T.~R., {et~al.} 2005, \mnras, 359, 1165

\bibitem[{{Hughes} {et~al.}(1998){Hughes}, {Serjeant}, {Dunlop},
  {Rowan-Robinson}, {Blain}, {Mann}, {Ivison}, {Peacock}, {Efstathiou}, {Gear},
  {Oliver}, {Lawrence}, {Longair}, {Goldschmidt}, \& {Jenness}}]{Hughes1998}
{Hughes}, D.~H., {et~al.} 1998, \nat, 394, 241

\bibitem[{{Ilbert} {et~al.}(2009){Ilbert}, {Capak}, {Salvato}, {Aussel},
  {McCracken}, {Sanders}, {Scoville}, {Kartaltepe}, {Arnouts}, {Le Floc'h},
  {Mobasher}, {Taniguchi}, {Lamareille}, {Leauthaud}, {Sasaki}, {Thompson},
  {Zamojski}, {Zamorani}, {Bardelli}, {Bolzonella}, {Bongiorno}, {Brusa},
  {Caputi}, {Carollo}, {Contini}, {Cook}, {Coppa}, {Cucciati}, {de la Torre},
  {de Ravel}, {Franzetti}, {Garilli}, {Hasinger}, {Iovino}, {Kampczyk},
  {Kneib}, {Knobel}, {Kovac}, {Le Borgne}, {Le Brun}, {F{\`e}vre}, {Lilly},
  {Looper}, {Maier}, {Mainieri}, {Mellier}, {Mignoli}, {Murayama}, {Pell{\`o}},
  {Peng}, {P{\'e}rez-Montero}, {Renzini}, {Ricciardelli}, {Schiminovich},
  {Scodeggio}, {Shioya}, {Silverman}, {Surace}, {Tanaka}, {Tasca}, {Tresse},
  {Vergani}, \& {Zucca}}]{Ilbert2009}
{Ilbert}, O., {et~al.} 2009, \apj, 690, 1236

\bibitem[{{Iono} {et~al.}(2006){Iono}, {Peck}, {Pope}, {Borys}, {Scott},
  {Wilner}, {Gurwell}, {Ho}, {Yun}, {Matsushita}, {Petitpas}, {Dunlop},
  {Elvis}, {Blain}, \& {Le Floc'h}}]{Iono2006}
{Iono}, D., {et~al.} 2006, \apjl, 640, L1

\bibitem[{{Ivison} {et~al.}(2007){Ivison}, {Greve}, {Dunlop}, {Peacock},
  {Egami}, {Smail}, {Ibar}, {van Kampen}, {Aretxaga}, {Babbedge}, {Biggs},
  {Blain}, {Chapman}, {Clements}, {Coppin}, {Farrah}, {Halpern}, {Hughes},
  {Jarvis}, {Jenness}, {Jones}, {Mortier}, {Oliver}, {Papovich},
  {P{\'e}rez-Gonz{\'a}lez}, {Pope}, {Rawlings}, {Rieke}, {Rowan-Robinson},
  {Savage}, {Scott}, {Seigar}, {Serjeant}, {Simpson}, {Stevens}, {Vaccari},
  {Wagg}, \& {Willott}}]{Ivison2007}
{Ivison}, R.~J., {et~al.} 2007, \mnras, 380, 199

\bibitem[{{Kennicutt}(1998)}]{Kennicutt1998}
{Kennicutt}, Jr., R.~C. 1998, \araa, 36, 189

\bibitem[{{Knudsen} {et~al.}(2010){Knudsen}, {Kneib}, {Richard}, {Petitpas}, \&
  {Egami}}]{Knudsen2010}
{Knudsen}, K.~K., {Kneib}, J., {Richard}, J., {Petitpas}, G., \& {Egami}, E.
  2010, \apj, 709, 210


\bibitem[{{Le Floc'h} {et~al.}(2005){Le Floc'h}, {Papovich}, {Dole}, {Bell},
  {Lagache}, {Rieke}, {Egami}, {P{\'e}rez-Gonz{\'a}lez}, {Alonso-Herrero},
  {Rieke}, {Blaylock}, {Engelbracht}, {Gordon}, {Hines}, {Misselt}, {Morrison},
  \& {Mould}}]{LeFloch2005}
{Le Floc'h}, E., {et~al.} 2005, \apj, 632, 169


\bibitem[{{Micha{\l}owski} {et~al.}(2010){Micha{\l}owski}, {Watson}, \&
  {Hjorth}}]{Michalowski2010}
{Micha{\l}owski}, M.~J., {Watson}, D., \& {Hjorth}, J. 2010, \apj, 712, 942

\bibitem[{{Mobasher} {et~al.}(2007){Mobasher}, {Capak}, {Scoville}, {Dahlen},
  {Salvato}, {Aussel}, {Thompson}, {Feldmann}, {Tasca}, {Lefevre}, {Lilly},
  {Carollo}, {Kartaltepe}, {McCracken}, {Mould}, {Renzini}, {Sanders},
  {Shopbell}, {Taniguchi}, {Ajiki}, {Shioya}, {Contini}, {Giavalisco},
  {Ilbert}, {Iovino}, {Le Brun}, {Mainieri}, {Mignoli}, \&
  {Scodeggio}}]{Mobasher2007}
{Mobasher}, B., {et~al.} 2007, \apjs, 172, 117

\bibitem[{{Polletta} {et~al.}(2007){Polletta}, {Tajer}, {Maraschi},
  {Trinchieri}, {Lonsdale}, {Chiappetti}, {Andreon}, {Pierre}, {Le F{\`e}vre},
  {Zamorani}, {Maccagni}, {Garcet}, {Surdej}, {Franceschini}, {Alloin},
  {Shupe}, {Surace}, {Fang}, {Rowan-Robinson}, {Smith}, \&
  {Tresse}}]{Polletta2007}
{Polletta}, M., {et~al.} 2007, \apj, 663, 81

\bibitem[{{Schinnerer} {et~al.}(2007){Schinnerer}, {Smol{\v c}i{\'c}},
  {Carilli}, {Bondi}, {Ciliegi}, {Jahnke}, {Scoville}, {Aussel}, {Bertoldi},
  {Blain}, {Impey}, {Koekemoer}, {Le Fevre}, \& {Urry}}]{Schinnerer2007}
{Schinnerer}, E., {et~al.} 2007, \apjs, 172, 46

\bibitem[{{Schinnerer} {et~al.}(2008){Schinnerer}, {Carilli}, {Capak},
  {Martinez-Sansigre}, {Scoville}, {Smol{\v c}i{\'c}}, {Taniguchi}, {Yun},
  {Bertoldi}, {Le Fevre}, \& {de Ravel}}]{Schinnerer2008}
---. 2008, \apjl, 689, L5

\bibitem[{{Scoville} {et~al.}(1993){Scoville}, {Carlstrom}, {Chandler},
  {Phillips}, {Scott}, {Tilanus}, \& {Wang}}]{Scoville1993}
{Scoville}, N.~Z., {Carlstrom}, J.~E., {Chandler}, C.~J., {Phillips}, J.~A.,
  {Scott}, S.~L., {Tilanus}, R.~P.~J., \& {Wang}, Z. 1993, \pasp, 105, 1482

\bibitem[{{Smail} {et~al.}(1997){Smail}, {Ivison}, \& {Blain}}]{Smail1997}
{Smail}, I., {Ivison}, R.~J., \& {Blain}, A.~W. 1997, \apjl, 490, L5+


\bibitem[{{Viero} {et~al.}(2009){Viero}, {Ade}, {Bock}, {Chapin}, {Devlin},
  {Griffin}, {Gundersen}, {Halpern}, {Hargrave}, {Hughes}, {Klein},
  {MacTavish}, {Marsden}, {Martin}, {Mauskopf}, {Moncelsi}, {Negrello},
  {Netterfield}, {Olmi}, {Pascale}, {Patanchon}, {Rex}, {Scott}, {Semisch},
  {Thomas}, {Truch}, {Tucker}, {Tucker}, \& {Wiebe}}]{Viero2009}
{Viero}, M.~P., {et~al.} 2009, \apj, 707, 1766

\bibitem[{{Wang} {et~al.}(2007){Wang}, {Cowie}, {van Saders}, {Barger}, \&
  {Williams}}]{Wang2007}
{Wang}, W., {Cowie}, L.~L., {van Saders}, J., {Barger}, A.~J., \& {Williams},
  J.~P. 2007, \apjl, 670, L89

\bibitem[{{Younger} {et~al.}(2007){Younger}, {Fazio}, {Huang}, {Yun}, {Wilson},
  {Ashby}, {Gurwell}, {Lai}, {Peck}, {Petitpas}, {Wilner}, {Iono}, {Kohno},
  {Kawabe}, {Hughes}, {Aretxaga}, {Webb}, {Mart{\'{\i}}nez-Sansigre}, {Kim},
  {Scott}, {Austermann}, {Perera}, {Lowenthal}, {Schinnerer}, \& {Smol{\v
  c}i{\'c}}}]{Younger2007}
{Younger}, J.~D., {et~al.} 2007, \apj, 671, 1531

\bibitem[{{Younger} {et~al.}(2008){Younger}, {Fazio}, {Wilner}, {Ashby},
  {Blundell}, {Gurwell}, {Huang}, {Iono}, {Peck}, {Petitpas}, {Scott},
  {Wilson}, \& {Yun}}]{Younger2008}
---. 2008, \apj, 688, 59

\bibitem[{{Younger} {et~al.}(2009){Younger}, {Fazio}, {Huang}, {Yun}, {Wilson},
  {Ashby}, {Gurwell}, {Peck}, {Petitpas}, {Wilner}, {Hughes}, {Aretxaga},
  {Kim}, {Scott}, {Austermann}, {Perera}, \& {Lowenthal}}]{Younger2009}
---. 2009, \apj, 704, 803

\bibitem[{{Younger} {et~al.}(2010){Younger}, {Fazio}, {Ashby}, {Civano},
  {Elvis}, {Gurwell}, {Huang}, {Iono}, {Peck}, {Petitpas}, {Scott}, {Wilner},
  {Wilson}, \& {Yun}}]{Younger2010}
---. 2010, ArXiv:1003.4264

\end{thebibliography}

\clearpage

\end{document}